\documentclass[preprint,aps,floats,nofootinbib,amssymb]{revtex4}
\usepackage{epsf,epsfig}

\usepackage{graphicx}
\usepackage{mathrsfs}

\newcommand{\be}{\begin{equation}}
\newcommand{\ee}{\end{equation}}
\newcommand{\bea}{\begin{eqnarray}}
\newcommand{\eea}{\end{eqnarray}}

\newcommand{\ba}{\begin{array}}
\newcommand{\ea}{\end{array}}
\newcommand{\bi}{\begin{itemize}}
\newcommand{\ei}{\end{itemize}}
\newcommand{\mev}{\; \textrm{MeV}}
\newcommand{\gev}{\; \textrm{GeV}}
\newcommand{\tev}{\; \textrm{TeV}}
\newcommand{\noi}{\noindent}
\newcommand{\tb}{\bar{t}}
\newcommand{\bb}{\bar{b}}
\newcommand{\thsd}{\times 10^{3}}
\renewcommand{\slash}{\displaystyle{\not}}

\begin{document}

\title{\vspace*{.75in}
Triple-Top Signal of New Physics at the LHC}

\author{
Vernon Barger$^1$,
Wai-Yee Keung$^2$,
Brian Yencho$^1$
}

\affiliation{
$^1$Department of Physics, University of Wisconsin, Madison, WI 53706 \\
$^2$Department of Physics, University of Illinois, Chicago, IL 60607\\
\vspace*{.5in}}

\thispagestyle{empty}

\begin{abstract}
\noindent We present leading-order (LO) cross sections for the production of three top quarks ($tt\tb$,$t\tb\tb$) at the Large Hadron Collider (LHC).  We find a total LO cross section for triple-top production in the Standard Model of $\sigma \approx 2 $ fb at $\sqrt{s}=14$ TeV and we give examples of two new physics models which have a significant enhancement to this channel. In the Minimal Supersymmetric Standard Model (MSSM), there are regions of parameter space where the decays of gluino pairs into final states including three tops has a cross section $\sigma \approx 41$ fb.  In a leptophobic $Z'$ model featuring right-handed couplings of the $u$-quark to the top, we find  $\sigma \approx 28$ fb.  With efficient identification and reconstruction of the top quarks, the triple-top signal could potentially provide evidence for new physics at the LHC.
\end{abstract}
\maketitle

\section{Introduction}

The top quark, with a measured mass of $m_{t} = 171.3 \pm 1.1 \pm 1.2 \gev$ \citep{pdg2009}, is the most massive of the elementary particles.  It is this distinction that makes the top quark so interesting to phenomenology as it is typically the most closely related to proposals of new physics beyond the Standard Model (SM).

As a leading potential mode of new physics discovery, top-pair production has been studied extensively for both the Fermilab Tevatron and for the Large Hadron Collider (LHC) (see \cite{han2008} and \cite{morrissey2009} and references therein for a thorough review of LHC top physics).  Single top production is also of interest, particularly for the $V_{tb}$ mixing element of the SM, and was recently observed at the Tevatron \cite{li2009}.  In this paper, we discuss the observation of triple-top events at the LHC and how this channel can be significantly affected by new physics.

In Section \ref{sec:sm}, we evaluate the leading-order (LO) SM production cross sections at the LHC for up to four tops while highlighting the triple-top signal.  We then describe two new physics models in Section \ref{sec:np} and discuss the results in Section \ref{sec:disc}.

\section{Standard Model Top Production}
\label{sec:sm}

\begin{figure}[tb]
\centering
\includegraphics[width=0.35\textwidth]{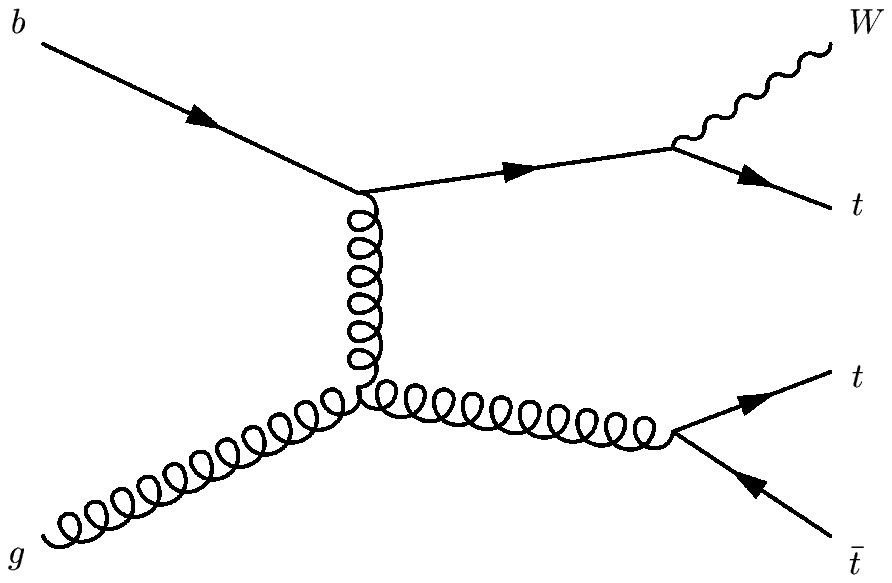} \qquad
\includegraphics[width=0.35\textwidth]{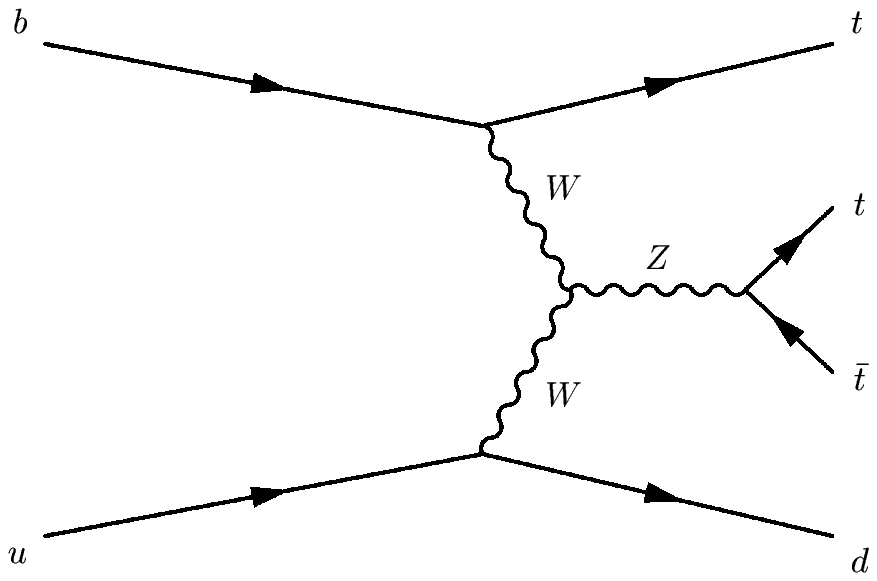}
\\ (a) \qquad \qquad \qquad \qquad  \qquad  \qquad  \qquad \qquad (b) \\
 \qquad \\
\includegraphics[width=0.35\textwidth]{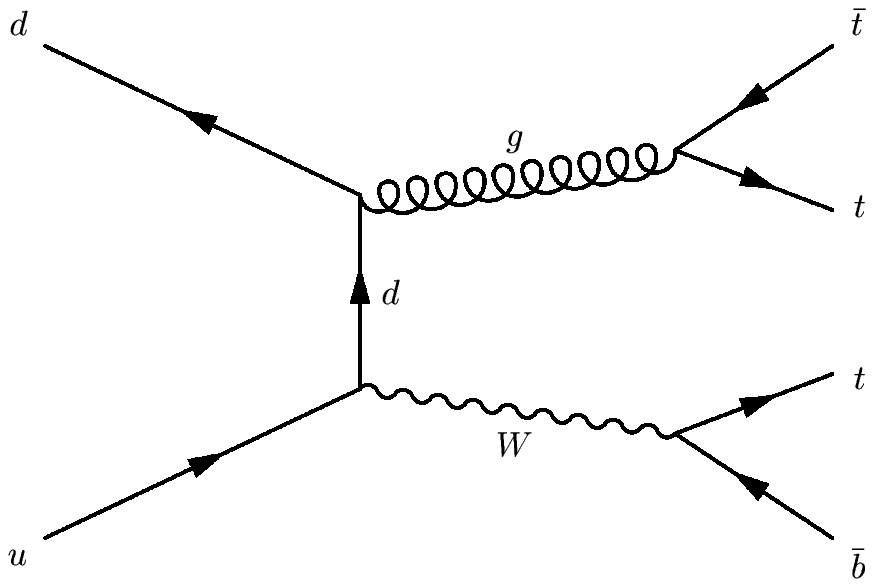}
\\ (c) 
\caption{Leading diagrams contributing to SM triple-top production at the LHC with $\sqrt{s}=14\tev$ corresponding to (a) $3t + W^{\pm}$ (b) $3t + {\rm jets}$ (c) $3t + b$.  The total number of $\textit{unique}$ diagrams (not summing over quark flavors or distinguishing between order of initial state partons) contributing to each process are (a) 236 (b) 144 and (c) 72.  All of these are included in our calculations.}
\label{fig:top}
\end{figure}

In the Standard Model, triple-top production, like single-top production, comes from three distinct processes at LO: $pp \to 3t+W^{\pm}, \; 3t+b, \; 3t+{\rm jets}$, where $t$ generically denotes top and anti-top quarks.  The typically dominant diagram from each of these processes is given in Figures \ref{fig:top}(a)-(c) and a summary of the total cross sections for each, calculated from all contributing diagrams, is given in Table \ref{tab:cSM} at three LHC center-of-mass energies, $\sqrt{s}=7,10,$ and 14 TeV. We calculate all cross sections to an accuracy of $1\%$ using MadGraph/MadEvent version 4.4.24 \cite{maltoni2002,alwall2007}, allowing the renormalization and factorization scales to vary with the process and using the default values on all cuts except $\eta_{\rm max}$ of the jets, which we set to 4.0, assuming forward-jet tagging at ATLAS and CMS \cite{aad2009,adolphi2008}.  We choose $m_{t}=171.4\gev$ and a Higgs boson mass of $m_{h}=130 \gev$.

\begin{figure}[tb]
\centering
\includegraphics[width=0.75\textwidth]{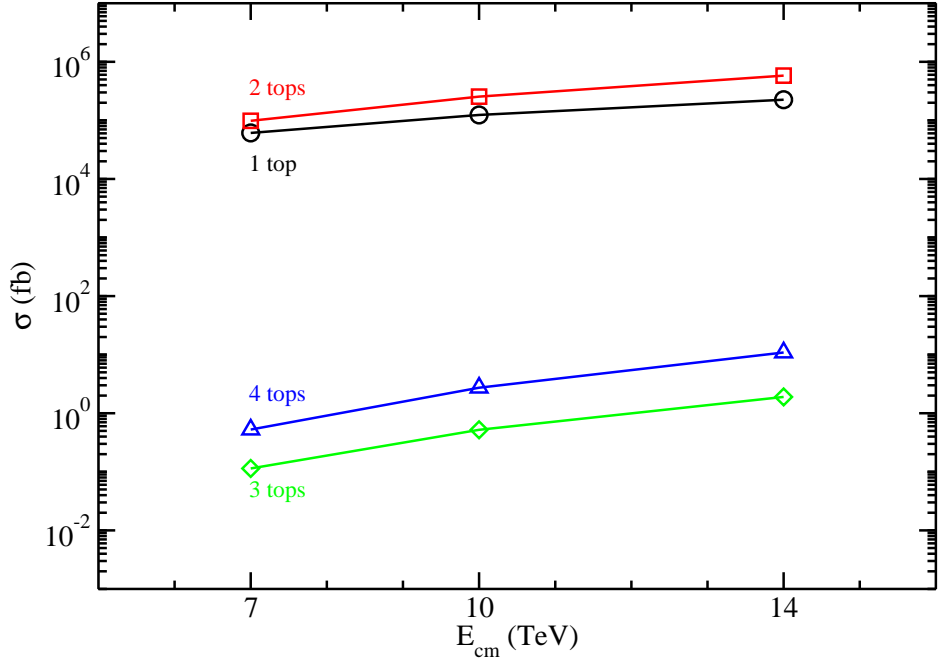}
\caption{Inclusive LO cross sections for multi-top production in the Standard Model with $m_{H}=130\gev$ for three LHC center-of-mass energies.  Single- and triple-top curves represent the sum over all contributing final states, which are given in Table \ref{tab:cSM}.}
\label{fig:cSM}
\end{figure}

Unlike top-pair and even four-top production\footnote{The production of four-heavy-quark final states at hadron colliders, including $t\tb t\tb$ at supercollider energies, was first calculated in \cite{barger1991}.}, which are dominated at the LHC by the strong coupling $gg \to t \tb$ process, the production of an odd number of tops requires a $W t b$ vertex in every diagram and often involves a $b$-quark in the initial state of the hard process, both of which result in a significant suppression compared to the strong processes.  Table \ref{tab:cSM} compares the total LO cross sections for 1,2,3, and 4 top production for the three center-of-mass energies. Figure \ref{fig:cSM} plots this information while summing over all possible channels and, in the case of single- and triple-tops, the contributions from both $t$/$\bar{t}$ and $t t \bar{t}$/$t \bar{t} \bar{t}$, respectively.  At the design LHC energy of $\sqrt{s}=14\tev$, the triple-top cross section, with $\sigma=1.9$ fb, is five orders of magnitude less than that of top-pair production, the dominant mode of top creation at the LHC.  This relatively low SM production is precisely what makes three tops an interesting channel for investigating new physics, as there may be possibilities for notable enhancements above the SM.

\begin{table}[tb]
\centering
\begin{tabular}{|l|c|c|c|} \hline \hline
\multicolumn{4}{|c|}{Inclusive Cross Sections at the LHC (fb)} \\ \hline \hline
Process & $\sqrt{s} = 7\tev$ & $10\tev$ & $14\tev$ \\ 
\hline \hline
$pp \to 1 t $ (Total)  & 66$\thsd$  & 138$\thsd$ & 258$\thsd$ \\ 
\hline \hline
\qquad $pp \to t+\ldots$       &  42$\thsd$ &  85$\thsd$ & 154$\thsd$ \\ 
\qquad \qquad $pp \to t W^{-}$ & 5.4$\thsd$ &  14$\thsd$ &  32$\thsd$ \\
\qquad \qquad $pp \to t j$     &  35$\thsd$ &  68$\thsd$ & 117$\thsd$ \\
\qquad \qquad $pp \to t \bb$   & 1.8$\thsd$ & 3.1$\thsd$ & 4.8$\thsd$ \\
\qquad $pp \to \tb+\ldots$       &  24$\thsd$ &  53$\thsd$ & 104$\thsd$ \\ 
\qquad \qquad $pp \to \tb W^{+}$ & 5.4$\thsd$ &  14$\thsd$ &  32$\thsd$ \\
\qquad \qquad $pp \to \tb j$     &  18$\thsd$ &  37$\thsd$ &  69$\thsd$ \\
\qquad \qquad $pp \to \tb b$     &0.99$\thsd$ & 1.8$\thsd$ & 3.0$\thsd$ \\
\hline \hline
$pp \to t\tb$ & 98$\thsd$ & 255$\thsd$ & 581$\thsd$ \\ 
\hline \hline
$pp \to 3 t $ (Total) & 0.11 & 0.52 & 1.9 \\ 
\hline \hline
\qquad $pp \to tt\tb+\ldots$       & 0.072 & 0.31  & 1.1  \\ 
\qquad \qquad $pp \to tt\tb W^{-}$ & 0.027 & 0.16  & 0.69 \\
\qquad \qquad $pp \to tt\tb j$     & 0.024 & 0.091 & 0.27 \\
\qquad \qquad $pp \to tt\tb \bb$   & 0.021 & 0.054 & 0.11 \\
\qquad $pp \to t\tb\tb+\ldots$       & 0.042 & 0.21  & 0.83  \\ 
\qquad \qquad $pp \to t\tb\tb W^{+}$ & 0.028 & 0.16  & 0.68  \\
\qquad \qquad $pp \to t\tb\tb j$     & 0.008 & 0.033 & 0.10  \\
\qquad \qquad $pp \to t\tb\tb b$     & 0.0065& 0.019 & 0.045 \\
\hline \hline
$pp \to tt \tb\tb$ & 0.53 & 2.7 & 11 \\
\hline \hline
\end{tabular}
\caption{LO inclusive cross sections for multi-top production in the SM for three different center-of-mass energies at the LHC.}
\label{tab:cSM}
\end{table}

\section{New Physics Models}
\label{sec:np}

One of the most studied and believed by many to be the most likely candidate of physics beyond the Standard Model is supersymmetry.  It successfully stabilizes the quantum corrections to the Higgs mass, provides a suitable dark matter candidate, is consistent with gauge coupling unification, and is characterized by a space-time symmetry that is a natural extension of the Poincar$\acute{\rm e}$ group.  The generic softly-broken supersymmetric model with the least number of new particles is the Minimal Supersymmetric Standard Model (MSSM).  As the MSSM makes no assumptions on the precise method of supersymmetry breaking and instead includes all allowed soft SUSY-breaking terms, many of the parameters defining the masses, mixings, and couplings of the new particles are left undetermined, introducing over 100 arbitrary parameters.

The phenomenology of the MSSM is dependent on the choice of these parameters.  Specifying the method of SUSY breaking, or choosing parameters with relationships motivated by a particular preference, makes the problem of quantifying MSSM expectations more tractable.  The ``Snowmass Points and Slopes'' (SPS) are a collection of benchmarks derived from different SUSY-breaking methods that all provide viable, and yet markedly different, phenomenological behavior \cite{allanach2002}.

The parameter point of interest to the current study is SPS 2.  This is in the so called ``focus point'' region of SUSY parameter space \cite{fp} and is defined in the mSUGRA model by the point:

\be
m_{0} = 1450 \gev,\quad  m_{1/2} = 300 \gev,\quad  A_{0} = 0,\quad \tan\beta = 10,\quad \mu > 0
\ee

\noi where $m_{0}$ is the universal scalar mass, $m_{1/2}$ is the universal gaugino mass, $A_{0}$ is the common trilinear coupling, $\tan\beta$ is the ratio of the Higgs VEVs, and $\mu$ is the Higgsino mass parameter, all of which are evaluated at $M_{\rm GUT} \sim 10^{16} \gev$ and run down with the renormalization group equations to determine all the low-scale parameters of the theory.  A feature of this benchmark point is that the gluino mass, $m_{\tilde{g}}=782 \gev$, is less that that of the squarks.  Gluino decays to quark-squark pairs ($\tilde{g} \to  q \bar{\tilde{q}}_{L,R}, \; \bar{q} \tilde{q}_{L,R}$) are kinematically inaccessible, opening up three-body decays via virtual squarks:

\bea
\tilde{g} &\to& \chi^{0}_{i} q \bar{q} \nonumber \\
\tilde{g} &\to& \chi^{\pm}_{i} q \bar{q'}
\eea

\noi These decays can result in both single- and double-top final states with considerable branching fractions of $\approx 10\%-20\%$ when summing over all neutralinos and charginos, an enhancement that has been studied for similar regions of parameter space \cite{baer1990,barger1993,acharya2009}.  These branching fractions are given in Table \ref{tab:gluino}.

\begin{figure}[t]
\centering
\includegraphics[width=0.35\textwidth]{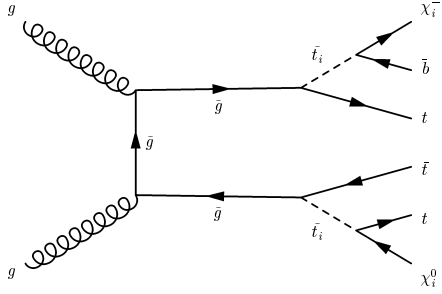}
\caption{Leading diagram contributing to triple-top production in the focus point region of the MSSM through the decay of two gluinos.  There are a total of 8 $\textit{unique}$ diagrams (not summing over quark flavors or distinguishing between the order of the initial state partons) contributing to gluino pair production.  We include all of these in our calculations.}
\label{fig:mssm}
\end{figure}

Because gluinos are strongly interacting, the production of gluino pairs at the LHC will be very large, typically several hundred fb.  With one gluino decaying to a single top via $\tilde{g} \to t \bar{b} \chi^{-}_{i}$ or $\; \tb b \chi^{+}_{i}$ and the other to a top pair through $\tilde{g} \to \chi^{0}_{i} t \bar{t}$, the rate of triple-top events can therefore be quite large in the MSSM in this region of parameter space.  The diagram for the leading contribution is given in Figure \ref{fig:mssm}.

The major limitations to the rate of triple-top production in the SM, as discussed in Section \ref{sec:sm}, are the $W t b$ vertex and the initial state $b$-quark needed to create the single top (though the latter is not involved for $pp\to tt\tb \bb$).  Models that can produce this single top with greater ease should therefore have a larger triple-top signal as well.  An example of one such model is the addition of a $U(1)'$ symmetry in which a leptophobic $Z'$ couples the top directly to the $u$-quark.  The additional interaction terms of the Lagrangian are given by

\be
\mathscr{L} \supset (g_{\rm X} Z_{\mu}' \bar{u} \gamma^{\mu} P_{R} t + h.c.) + \epsilon_{U} g_{\rm X} Z_{\mu}' \bar{u}_{i} \gamma^{\mu} P_{R} u_{i}
\ee

\begin{figure}[tb]
\centering
\includegraphics[width=0.35\textwidth]{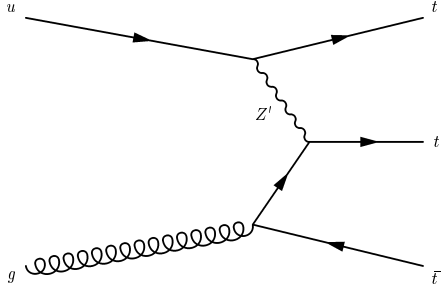}
\caption{Leading diagram contributing to triple-top production in a leptophobic $Z'$ in which the top quark can couple directly to a $u$-quark.  There are a total of 16 $\textit{unique}$ diagrams (not distinguishing between the order of  the initial state partons) involved in this process.  We include all of these in our calculations.}
\label{fig:zp}
\end{figure}

\noi where $\epsilon < 1$ and the diagonal term is summed over generations.  This model was discussed in \cite{jung2009} as a candidate for describing the observed $2 \sigma$ deviation from the SM prediction of forward-backward asymmetry in the top-pair signal at the CDF detector of the Tevatron collider \cite{aaltonen2008}.  The small diagonal couplings characterized by the parameter $\epsilon_{U}$ exist only to escape bounds on like-sign top quark events from the decay of two $Z'$s by forcing the dominant decay $Z' \to u \bar{u}$.  This study found the best match to the asymmetry and to the invariant mass distribution of the top pair with $M_{Z'}=160\gev$ and $\alpha_{X}=0.024$, with any small $\epsilon_{U}\ne 0$ giving comparable results.  We take these values and choose $\epsilon_{U}=0.1$.  The dominant process for triple-top production in this model (for small $\epsilon_{U}$) is the t-channel exchange of the $Z'$, shown in Figure \ref{fig:zp}.  This diagram illustrates the unique topology of these events in this $Z'$ model, with the three tops produced at LO.

\begin{table}[tb]
\centering
\begin{tabular}{|l|c|} \hline \hline
\multicolumn{2}{|c|}{Gluino Branching Fractions} \\ \hline \hline
$\tilde{g} \to 1 t +\ldots$    &  0.21 \\ \hline
\qquad $\tilde{g} \to t\bb \chi_{1}^{-}$    & 0.080 \\ 
\qquad $\tilde{g} \to \tb b\chi_{1}^{+}$    & 0.080 \\ 
\qquad $\tilde{g} \to t\bb \chi_{2}^{-}$    & 0.024 \\ 
\qquad $\tilde{g} \to \tb b\chi_{2}^{+}$    & 0.024 \\ \hline
$\tilde{g} \to 2 t +\ldots$    &  0.11 \\ \hline
\qquad $\tilde{g} \to t\tb \chi_{1}^{0}$    &  0.099 \\ 
\qquad $\tilde{g} \to t\tb \chi_{2}^{0}$    &  0.012 \\ 
\qquad $\tilde{g} \to t\tb \chi_{3}^{0}$    &  0\\ 
\qquad $\tilde{g} \to t\tb \chi_{4}^{0}$    &  0\\ 
\hline \hline
\end{tabular}
\caption{Branching fractions for gluino decay modes to tops in the focus point region of the MSSM (SPS 2).  Here the gluino has a mass of $m_{\tilde{g}}=782\gev$ and a total decay width of $\Gamma_{\tilde{g}}=2.6 \mev$.}
\label{tab:gluino}
\end{table}

\section{Results and Conclusions}
\label{sec:disc}

\begin{figure}[tb]
\centering
\includegraphics[width=0.45\textwidth]{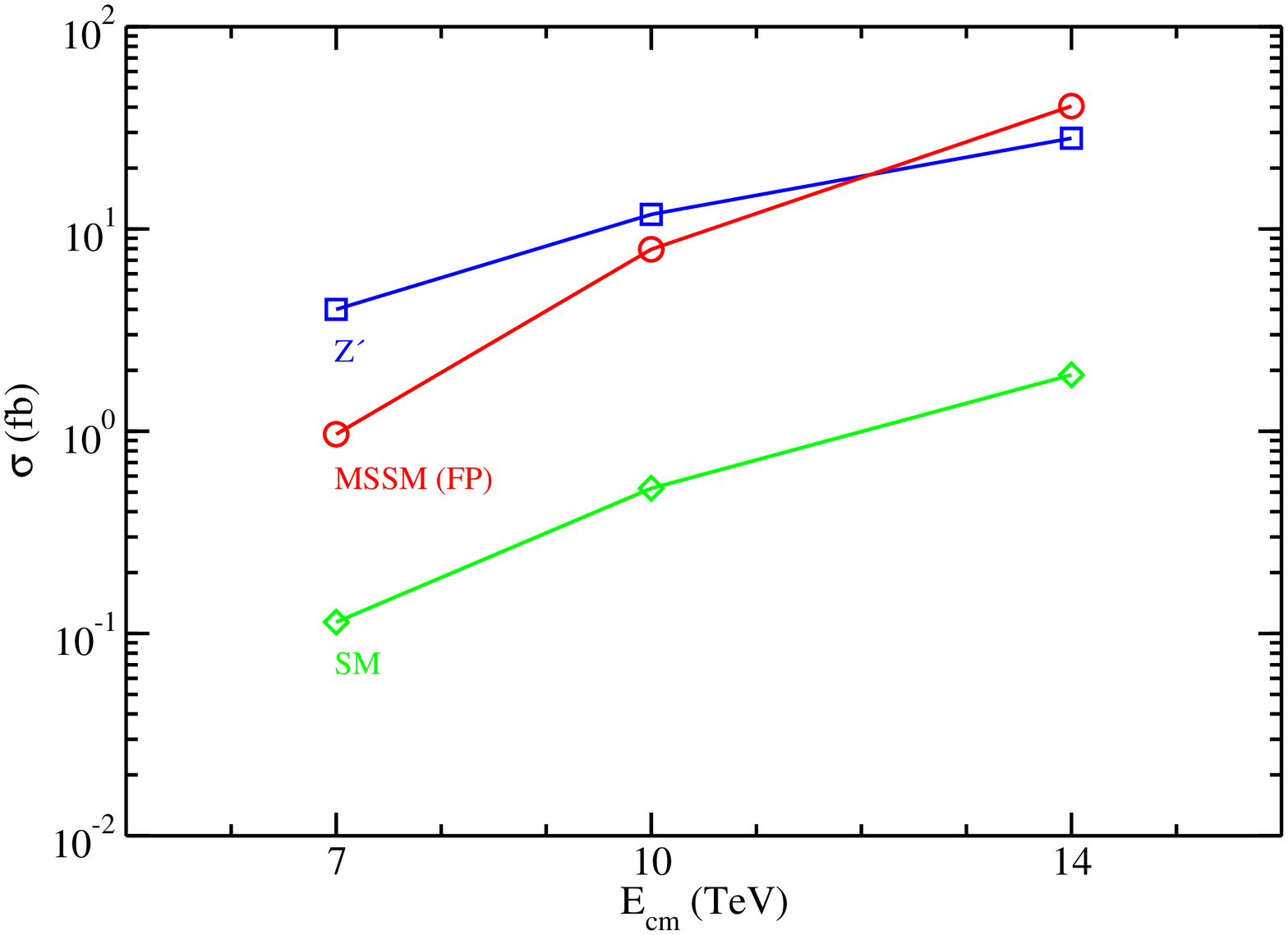} \qquad
\includegraphics[width=0.45\textwidth]{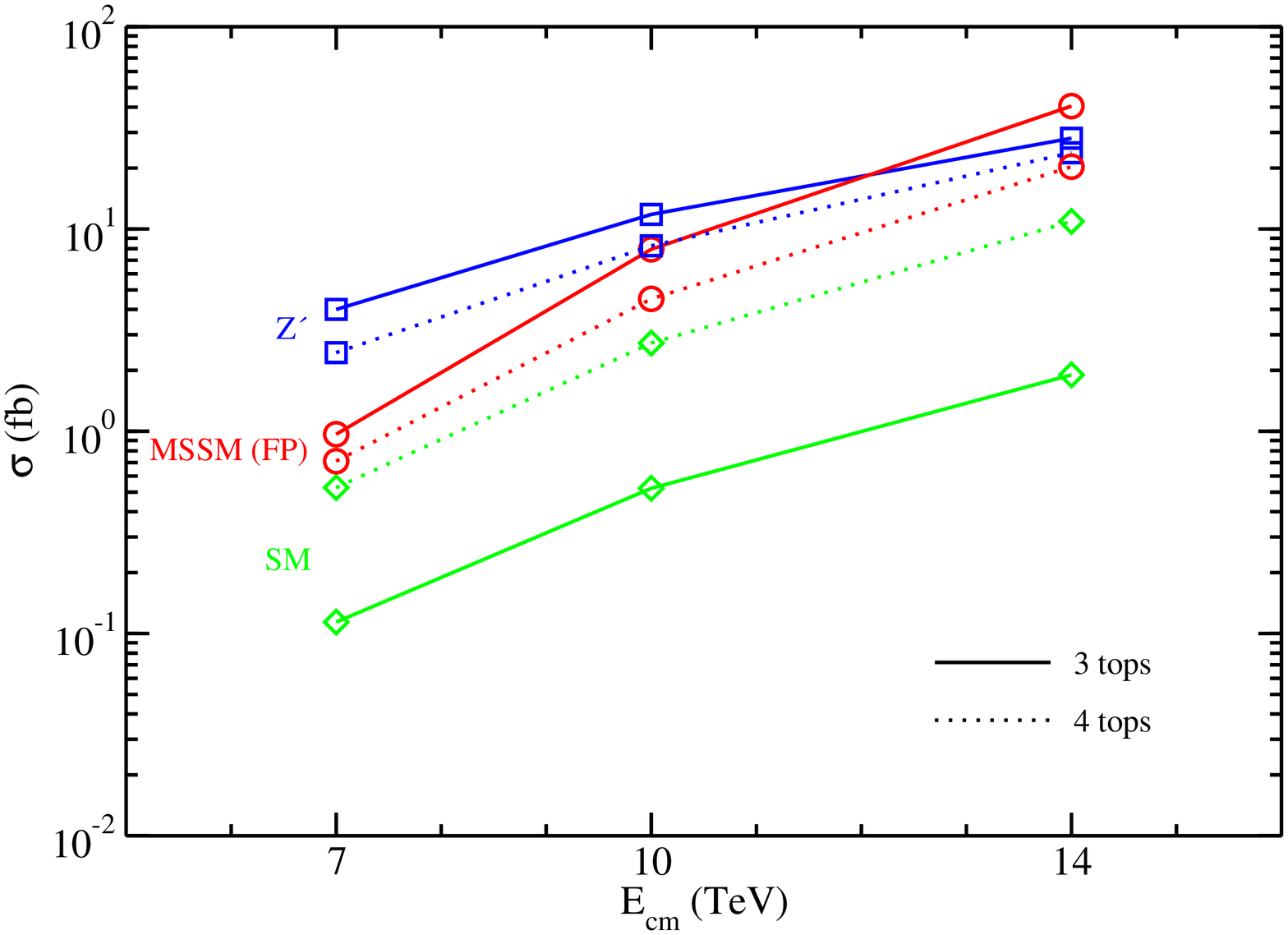} \\
(a) \hspace{150pt} (b)
\caption{(a) Inclusive LO cross sections for triple-top production at three LHC center-of-mass energies in the Standard Model, the $Z'$ model, and the focus point region of the MSSM through the decay of two gluinos.  Each curve represents the sum over all contributing final states, which are given in Tables \ref{tab:cSM} and \ref{tab:tritop}.  In (b), these same curves are plotted (solid lines) along with the corresponding four-top cross sections of each model (dotted lines), which are similarly summed over all contributing final states.}
\label{fig:tritop}
\end{figure}

The LO triple-top production cross sections for the two new physics models discussed are calculated with MadGraph according to the prescriptions in Section \ref{sec:sm}.  The $Z'$ model is implemented with the ``usermod'' format and the cross section for the three-top final state is found directly.  For the MSSM we use the standard MadGraph MSSM implementation and the parameter card for SPS 2 from the MadGraph webpage\footnote{It should be noted that while the current value of the top mass, $m_{t}=171.4\gev$, was used for the SM and $Z'$ model calculations, the MSSM at SPS 2 uses $m_{t}=175 \gev$.}. We calculate the cross section for the on-shell gluino pair and then determine the triple-top cross section using the branching fractions in Table \ref{tab:gluino}. The results for three different LHC running energies are given in Table \ref{tab:tritop} and the sum of all processes for each model is plotted with the SM results in Figure \ref{fig:tritop}(a). 

\begin{table}[tb]
\centering
\begin{tabular}{|l|c|c|c|} \hline \hline
\multicolumn{4}{|c|}{Inclusive Cross Sections at the LHC (fb)} \\ \hline \hline
$pp \to 3 t $ (Total) & $\sqrt{s} = 7\tev$ & $10\tev$ & $14\tev$ \\ 
\hline \hline
$Z^{\prime}$ model & 4.0 & 12 & 28 \\ 
\hline \hline
\qquad $pp \to tt\tb$    & 3.8  & 11   & 26  \\ 
\qquad $pp \to t\tb\tb$  & 0.20 & 0.75 & 2.2  \\ 
\hline \hline
MSSM & 0.97 & 7.9 & 41 \\ 
\hline \hline
\qquad $pp \to tt\tb + \bb \chi_{i}^{-} \chi_{j}^{0}$    & 0.49 & 4.0 & 21 \\ 
\qquad $pp \to t\tb\tb + b \chi_{i}^{+} \chi_{j}^{0}$   & 0.49 & 4.0 & 21 \\ 
\hline \hline
\end{tabular}
\caption{LO inclusive cross sections for triple-top production at the LHC for three different center-of-mass energies.  In MSSM, the final state neutralinos and charginos have been summed over.}
\label{tab:tritop}
\end{table}

At each energy studied, the new physics models lie approximately an order of magnitude above the SM prediction.  Each model will also involve different topologies and kinematics which may allow them to be distinguished with this signal, given enough data.  The SM triple-top events will often be associated with an extra $W$-boson or extra tagged $b$-quark, while the MSSM events would feature large $\slash{E_{T}}$ due to the neutralinos escaping the detector.  The $Z'$ model would best be characterized by the absence of these additional states and may also be distinguished by the presense of a broad pseudo-rapidity distribution of one top due to the t-channel exchange of the $Z'$-boson\footnote{When the tops are ordered by $p_{T}$, we find that the low-$p_{T}$ $t$ (as opposed to the $\tb$) in the $tt\tb$ events occurs with $|\eta|>2$ in $\approx 60\%$ of the events in the $Z'$ model, compared to $\approx 30\%$ in the SM and $\approx 16\%$ in the focus point region of the MSSM.  This was based on an analysis of 10,000 unweighted events.}

However, despite the large increase in cross section at all energies and distinct topologies, the actual detection of the triple-top signal for the models discussed here would likely not be possible for the initial run of the LHC at 7 TeV center-of-mass energy and with 1 fb$^{-1}$ of integrated luminosity.  Indeed, the given BSM models may be discovered through other signals before triple-top events can even be identified.  For example, in the $Z'$ model, there are large  cross sections for same-sign top production at 7 TeV ($\sim 50$ pb at LO) that would likely lead to discovery, and the MSSM could possibly be inferred at these lower energies from generic missing energy searches.  If supersymmetry has eluded detection during this initial run, however, the triple-top signal could prove to be an important part of detection at 14 TeV.  Indeed, the branching fractions of light gluinos to multi-top final states can be made much larger than those presented here if the 1st and 2nd generation squarks are much more massive than those of the 3rd, providing a significant boost to the triple-top cross section.  The discovery potential of the MSSM using multi-top signals in such a scenario is discussed in detail in \cite{acharya2009}. 

In addition to providing evidence for physics beyond the SM, knowledge of the triple-top cross sections can also aid in the understanding of the underlying models.  In the $Z'$ model, while the two- and four-top production cross sections are essentially independent of the parameter $\epsilon_{U}$, every diagram contributing to the production of three tops has a linear dependence on this parameter and therefore the total cross section goes as $\epsilon_{U}^{2}$.  Knowledge of the triple-top signal in addition to the two- and four-top signals could therefore allow $\epsilon_{U}$ to be determined.  In the focus-point region of the MSSM, the relative sizes of the various multi-top signals are determined by the relative branching fractions of the gluinos to one- and two-top final states.  In certain scenarios, the ratios of these branching fractions can provide insight into the 3rd generation squark mass parameters and the composition of the lightest neutralino, while the overall rates and total effective mass distributions can constrain the gluino mass itself \cite{acharya2009}.

An important question to then consider is whether a BSM triple-top signal could be faked by SM four-top production with one top going unidentified.  In Figure~\ref{fig:tritop}(b), we compare the triple-top cross sections predicted by each model with the corresponding four-top cross sections calculated at LO.  Both the $Z'$ and MSSM triple-top cross sections lie well above the SM four-top cross sections at all energies.  This reduces the chance that a SM four-top signal could fake the larger BSM triple-top one.  Each model also lies above its own four-top cross sections, which are themselves enhanced from the SM prediction.  It should be noted, however, that these conclusions are not generic features of the models; slightly different values of $\epsilon_{U}$ and the gluino branching fractions can flip the relative importance of the three- and four-top channels.  It should be understood, therefore, that the triple-top signal is most valuable in its relation to top signals of other multiplicities.

Means of distinguishing between two-, three-, and four-top final states are then critical to properly utilizing these signals.  Direct reconstruction of every top quark in a multi-top event, after including detector cuts, background, and showering, can be extremely difficult  due to large combinatorics \cite{acharya2009}.  Other methods have been developed for tagging high-$p_{T}$ top quarks at the LHC by investigating the jet substructure of the decay products \cite{kaplan2008,barger2008,giurgiu2009,krohn2009,plehn2009}.  These studies have focused on top-pair signals but perhaps could find applicability to three- and four-top events as well.  While there are certainly practical challenges involved, the results presented here indicate that new physics may be inferred from the triple-top signal and therefore warrant further, more detailed simulations.

\section{Acknowledgements}

The authors would like to thank the referee for valuable comments that have contributed to quality of this manuscript and also Y. Gao, I. Lewis, and M. McCaskey for helpful discussions.  This work was supported in part by the U.S. Department of Energy under grants Nos.~DE-FG02-95ER40896 and DE-FG02-84ER40173 and in part by the Wisconsin Alumni Research Foundation.

\newpage
\bibliographystyle{h-physrev}
\bibliography{ms.bbl}

\begin{thebibliography}{100}

\bibitem{pdg2009}
  C. Amsler {\it et al.} (Particle Data Group),
  Physics Letters {\bf B667}, 1 (2008)
  and 2009 partial update for the 2010 edition

\bibitem{han2008}
  T.~Han,
  Int.\ J.\ Mod.\ Phys.\  A {\bf 23}, 4107 (2008)
  [arXiv:0804.3178 [hep-ph]].

\bibitem{morrissey2009}
  D.~E.~Morrissey, T.~Plehn and T.~M.~P.~Tait,
  arXiv:0912.3259 [hep-ph].

\bibitem{li2009}
  L.~Li  [D0 Collaboration],
  arXiv:0911.1150 [hep-ex].

\bibitem{maltoni2002}
  F.~Maltoni and T.~Stelzer,
  JHEP {\bf 0302} (2003) 027
  [arXiv:hep-ph/0208156].

\bibitem{alwall2007}
  J.~Alwall {\it et al.},
  JHEP {\bf 0709}, 028 (2007)
  [arXiv:0706.2334 [hep-ph]].

\bibitem{aad2009}
  G.~Aad {\it et al.}  [The ATLAS Collaboration],
  arXiv:0901.0512 [hep-ex].

\bibitem{adolphi2008}
  R.~Adolphi {\it et al.}  [CMS Collaboration],
  JINST {\bf 0803} (2008) S08004
  [JINST {\bf 3} (2008) S08004].

\bibitem{barger1991}
  V.~D.~Barger, A.~L.~Stange and R.~J.~N.~Phillips,
  Phys.\ Rev.\  D {\bf 44}, 1987 (1991).

\bibitem{allanach2002}
  B.~C.~Allanach {\it et al.},
in {\it Proc. of the APS/DPF/DPB Summer Study on the Future of Particle Physics (Snowmass 2001) } ed. N.~Graf,
  Eur.\ Phys.\ J.\  C {\bf 25}, 113 (2002)
  [arXiv:hep-ph/0202233].

\bibitem{fp}
  K.~L.~Chan, U.~Chattopadhyay and P.~Nath, 
  Phys.\ Rev.\  D  {\bf 58}, 096004 (1998)
  [arXiv:hep-ph/9710473]; 
  J.~L.~Feng, K.~T.~Matchev and T.~Moroi,
  Phys.\ Rev.\ Lett.\  {\bf 84}, 2322 (2000)
  [arXiv:hep-ph/9908309];
  H.~Baer, C.~H.~Chen, F.~Paige and X.~Tata,
  Phys.\ Rev.\  D  {\bf 52}, 2746 (1995)
  [arXiv:hep-ph/9503271]

\bibitem{baer1990}
  H.~Baer, X.~Tata, and J.~Woodside,
  Phys.\ Rev.\  D {\bf 41}, 906 (1990);
  D {\bf 42}, 1568 (1991);
  D {\bf 45}, 142 (1992)

\bibitem{barger1993}
  V.~Barger, R.~J.~N.~Phillips and A.~L.~Stange,
  Phys.\ Rev.\  D {\bf 45}, 1484 (1992)

\bibitem{acharya2009}
  B.~S.~Acharya, P.~Grajek, G.~L.~Kane, E.~Kuflik, K.~Suruliz and L.~T.~Wang,
  arXiv:0901.3367 [hep-ph].

\bibitem{jung2009}
  S.~Jung, H.~Murayama, A.~Pierce and J.~D.~Wells,
  arXiv:0907.4112 [hep-ph].

\bibitem{aaltonen2008}
  T.~Aaltonen {\it et al.}  [CDF Collaboration],
  Phys.\ Rev.\ Lett.\  {\bf 101}, 202001 (2008)
  [arXiv:0806.2472 [hep-ex]].

\bibitem{kaplan2008}
  D.~E.~Kaplan, K.~Rehermann, M.~D.~Schwartz and B.~Tweedie,
  Phys.\ Rev.\ Lett.\  {\bf 101}, 142001 (2008)
  [arXiv:0806.0848 [hep-ph]].

\bibitem{barger2008}
  V.~Barger, T.~Han and D.~G.~E.~Walker,
  Phys.\ Rev.\ Lett.\  {\bf 100} (2008) 031801
  [arXiv:hep-ph/0612016].

\bibitem{giurgiu2009}
  G.~Giurgiu  [for the CMS collaboration],
  arXiv:0909.4894 [hep-ex].

\bibitem{krohn2009}
  D.~Krohn, J.~Shelton and L.~T.~Wang,
  arXiv:0909.3855 [hep-ph].

\bibitem{plehn2009}
  T.~Plehn, G.~P.~Salam and M.~Spannowsky,
  arXiv:0910.5472 [hep-ph].

\bibitem{butterworth2007}
  J.~M.~Butterworth, J.~R.~Ellis and A.~R.~Raklev,
  JHEP {\bf 0705}, 033 (2007)

\end{thebibliography}
\newpage

\end{document}